\documentclass[preprint]{elsarticle}

  \usepackage{latexsym,bm,amsmath,amssymb,amsfonts}
  \usepackage{epsfig,graphics,graphicx}
  \usepackage{slashed}
  \usepackage{cite}
  \usepackage[latin1]{inputenc}
  \usepackage{mathrsfs}
\usepackage{mathcomp}

  \long\def\comment#1{ }

  \newcommand{\eqnum}[1]{Eq.~\eqref{#1}}

  \newcommand{\abar}{\bar{\alpha}_s}
  
  \newcommand{\del}{\partial}

  \newcommand{\mcal}{\mathcal}
  \newcommand{\rme}{{\rm e}}
  \newcommand{\rmd}{{\rm d}}   

  \newcommand{\Lam}{\Lambda_{{\rm QCD}}}
  
  \newcommand{\nn}{\nonumber\\}
   
  \newcommand{\order}[1]{\mcal{O}{(#1)}}

  \newcommand{\beq}{\begin{eqnarray}}
  \newcommand{\eeq}{\end{eqnarray}}
  
 \def\simge{\mathrel{%
   \rlap{\raise 0.511ex \hbox{$>$}}{\lower 0.511ex \hbox{$\sim$}}}}
\def\simle{\mathrel{
   \rlap{\raise 0.511ex \hbox{$<$}}{\lower 0.511ex \hbox{$\sim$}}}}

\vspace{1.5cm}

\begin{document}
\begin{frontmatter}

\title{BFKL and CCFM evolutions with saturation boundary}

\author{Emil Avsar, Edmond Iancu}\ead{Emil.Avsar@cea.fr, Edmond.Iancu@cea.fr}

\address{Institut de Physique Th\'eorique de Saclay,
 F-91191 Gif-sur-Yvette, France}

\date{\today}
\vspace{1.2cm}
\begin{abstract}
We perform numerical studies of the BFKL and CCFM equations for the
unintegrated gluon distribution supplemented with an absorptive boundary
which mimics saturation. For the BFKL equation, this procedure yields the
same results for the saturation momentum and the gluon distribution above
saturation as the non--linear BK equation, for both fixed and running
coupling, and for all the considered energies. This similarity goes
beyond expectations based on the correspondence with statistical physics,
which hold only for fixed coupling and asymptotically high energies. For
the CCFM equation, whose non--linear generalization is not known, our
method provides the first study of the approach towards saturation. We
find that, in the running--coupling case, the CCFM and BFKL predictions
for the energy dependence of the saturation momentum are identical within
our numerical accuracy. A similar saturation boundary could be easily
implemented in the CCFM--based Monte Carlo event generators, so like
CASCADE.
\end{abstract}

\end{frontmatter}

\section{Introduction}

The imminent high--energy experiments at LHC will considerably enlarge
the phase--space where the unitarity corrections to QCD interactions,
like gluon saturation and multiple scattering, are expected to be
important. Such corrections should in particular influence some `hard'
observables, like jet production at forward rapidities, whose theoretical
description lies within the realm of perturbative QCD. The jets to be
measured at LHC will carry relatively large transverse momenta $Q\ge
10$~GeV, but because of the high--energy kinematics, their description
may go beyond the standard pQCD formalism at high $Q^2$ --- the DGLAP
evolution \citep{DGLAP} of the parton distributions together with the
collinear factorization of the hadronic cross--sections. Rather, the
high--energy evolution, of the BFKL \citep{BFKL} or CCFM \citep{CCFM}
type, and the associated $k_T$--factorization scheme should prevail
whenever the energy logarithms $\ln s\sim Y$ are larger than the momentum
ones, $\ln Q^2$. Besides, this evolution is expected to be amended by
non--linear effects reflecting gluon saturation and multiple scattering,
whose theoretical description within pQCD has been given only recently
\citep{Balitsky:1995ub,Kovchegov:1999yj,JKLW,CGC,CGCreviews} and (in full
rigor) only in the leading--order approximation (see Refs.
\citep{Balitsky:2006wa,Kovchegov:2006vj} for recent extensions to running
coupling). Such effects can make themselves felt even at relatively large
momenta $Q$, well above the saturation momentum $Q_s$ (the characteristic
scale for the onset of unitarity corrections), via phenomena like
geometric scaling
\citep{Stasto:2000er,Iancu:2002tr,Mueller:2002zm,Triantafyllopoulos:2002nz,Munier:2003vc,
Marquet:2006jb, Gelis:2006bs}, which reflect the change in the
unintegrated gluon distribution at high $k_\perp\gg Q_s$ due to
saturation at low $k_\perp\lesssim Q_s$. The saturation scale $Q_s$
grows, roughly, like a power of the energy: $Q_s^2\sim s^\lambda$ with
$\lambda\simeq 0.25$ from fits to HERA data
\citep{Iancu:2003ge,Soyez:2007kg}, which are also supported by
next--to--leading order theoretical calculations
\citep{Triantafyllopoulos:2002nz}. For forward jet production in
proton--proton collisions at LHC, $Q_s$ is expected in the ballpark of 2
to 3 GeV. Besides, much higher values of $Q_s$ can be effectively reached
\citep{Iancu:2008kb} by focusing on `hot spots' (partons at large $x$ and
high $Q^2$, which develop their own gluon cascades and saturation
momentum) in some rare events, so like Mueller--Navelet jets.

In view of the above, it becomes important and urgent to provide
realistic, quantitative, predictions for the effects of saturation on
relatively hard ($Q^2\gg Q^2_s$) observables at LHC. Besides making the
whole perturbative approach fully justified, the restriction to
relatively hard momenta entails some important simplifications, which are
essential for the strategy that we shall propose in this Letter.

First, this implies that one can neglect the complex many--body
correlations which develop at saturation ($Q\lesssim Q_s$), but focus on
the gluon phase--space density (or `unintegrated gluon distribution')
alone. Hence, the standard $k_T$--factorization of the hadronic
cross--sections still applies, but with modified unintegrated gluon
distributions, which reflect saturation. This opens the possibility to
include the effects of saturation within Monte--Carlo event generators
based on $k_T$--factorization, so like CASCADE \citep{Jung:2000hk},
which relies on the CCFM evolution.

Second, this means that the saturation effects in the gluon distribution
and, in particular, the saturation momentum itself can be computed
without a detailed knowledge of the non--linear dynamics responsible for
unitarization. Rather, they are fully determined by the {\em linear}
evolution if the latter is supplemented with an absorptive boundary
condition at low momenta, whose position is energy--dependent (as this
mimics the saturation momentum) and is self--consistently determined by
the evolution \citep{Iancu:2002tr,Mueller:2002zm,Triantafyllopoulos:2002nz}
. This
property is both important and highly non--trivial. It is important as it
allows one to perform studies of saturation even for evolution equations
whose non--linear generalizations are not known, so like the CCFM
evolution, and also the BFKL evolution beyond the leading--order
approximation. It is moreover non--trivial since the high--energy
evolution is non--local in transverse momenta, hence the growth in the
gluon distribution at high momenta $k_\perp\gg Q_s$ could be well feed by
radiation from lower momenta $k_\perp\lesssim Q_s$. This is clearly the
case for the linear evolution with running coupling, in which the gluon
distribution grows faster in the infrared (where the coupling is
stronger) and then acts as a source for radiating gluons at higher
momenta.

Yet, at least for a {\em fixed coupling} and for {\em asymptotically high
energies}, it has been firmly established that the high--energy evolution
towards saturation is driven by the linear evolution in the dilute tail
of the gluon distribution at high transverse momenta $k_\perp\gg Q_s$.
The respective argument is based on a correspondence between high--energy
QCD and statistical physics \citep{Munier:2003vc,Iancu:2004es}, which
however fails to apply in the more realistic case of a {\em running
coupling} \citep{Dumitru:2007ew}. For that case, our subsequent numerical
results will provide the first unambiguous evidence that the evolution of
the saturation front in QCD is indeed driven by the linear part of the
evolution, at {\em all} energies, including the preasymptotic ones. Most
precisely, we shall find that the BFKL equation with saturation boundary
provides exactly the same results for the saturation momentum $Q_s(Y)$
and for the gluon distribution at $k_\perp\ge Q_s(Y)$ as the non--linear
Balitsky--Kovchegov (BK) equation \citep{Balitsky:1995ub,Kovchegov:1999yj}
, for
both fixed and running coupling, and for all the considered rapidities
$Y\le 120$ --- including lower values $Y\le 14$, as relevant for the
phenomenology at LHC.

But the BFKL equation and its non--linear, BK, extension, to be discussed
in Sect.~2, will merely serve as a playground to test our method for
numerically implementing the saturation boundary condition within a
generic linear evolution. Our main interest is rather in the CCFM
evolution, that we shall discuss in Sect.~3, and which stays at the basis
of Monte--Carlo event generators \citep{Jung:2000hk}. There are at least
two reasons why the CCFM evolution is a privileged tool in that respect.
First, it takes into account the quantum coherence between successive
emissions, leading to angular ordering in the parton cascades. This
allows for a more realistic description of the particle distribution in
the final state as compared to the BFKL evolution.  (The latter only
guarantees the correct treatment of inclusive quantities like the
unintegrated gluon distribution.) Second, the CCFM equation provides an
interpolation between BFKL dynamics at small $x$ and (approximate) DGLAP
dynamics at larger $x$. Of course, when discussing saturation we are {\em
a priori} interested in the small--$x$ region, and there is little doubt
that, for asymptotically high energies, the approach towards saturation
is more correctly described by the BFKL formalism. However, the energies
to be available at LHC are far from asymptotia, and a better treatment of
the transverse momentum ordering within the parton cascades, as implicit
in the CCFM formalism, is probably essential for most of the
``small--$x$'' phase--space to be experimentally accessible at LHC.

In this Letter, we shall limit ourselves to applications of the CCFM and
BFKL evolutions of the unintegrated gluon distribution, so it will be
meaningful to compare their results. To achieve a faster numerical
convergence, we shall solve the respective differential equations, rather
than building Monte--Carlo codes. But our prescription for implementing
the saturation boundary condition is straightforward to apply within a
Monte--Carlo event generator, so like CASCADE. Our present analysis can
be viewed as a first step towards building an event generator which
includes saturation \citep{wip}. The CCFM equation is considerably more
involved than the BFKL one (in particular, its solution involves one
additional variable: a maximum angle), so in order to be able to follow
this evolution up to relatively high rapidities we shall consider a
slightly simplified version of it. The simplification is obtained by
rewriting the CCFM equation in a `more inclusive form', i.e., by using
the virtual, Sudakov--like, terms to cancel some of the real gluon
emissions \citep{Andersson:1995ju,Salam:1999ft}. This procedure also
involves some kinematical approximations, which however are in the spirit
of the CCFM formalism\footnote{A more detailed discussion of the CCFM
equation and its various rewritings will be given in a longer publication
\citep{Avsar}.}. By solving the ensuing equation with saturation boundary
condition, we shall for the first time study the onset of unitarity
corrections within the CCFM evolution. One of our most interesting
results is that, in the running coupling case, the CCFM evolution in the
presence of saturation provides almost identical results (for the
saturation momentum and the gluon spectrum above $Q_s$) as the respective
BFKL evolution.

\section{BFKL evolution with absorptive boundary}

In this section we shall explain our method for effectively implementing
saturation within a unitarity--violating linear evolution on the example
of the BFKL equation \citep{BFKL}. This is interesting since the
corresponding non--linear equation which obeys unitarity is known as
well, the BK equation \citep{Balitsky:1995ub,Kovchegov:1999yj}, and thus
it can be used to test our method. Although all our numerical studies
will be performed in (transverse) momentum space (referring to the
unintegrated gluon distribution), it is more convenient to explain our
method in coordinate space. Then, the BK equation describes the
high--energy evolution of the scattering amplitude $T(Y,r)$ of a small
quark--antiquark dipole with transverse size $r$. We shall assume the
target to be infinite and homogeneous in transverse directions, so we can
ignore the impact--parameter dependence of the scattering amplitude and
average over angles. The corresponding equation reads
 \beq\label{BK}
{\partial \, T(Y,r)\over {\partial Y}}&=& \frac{\bar\alpha_s}{2\pi}\int
d^2{\bm z} {{r^2} \over {z^2 ({\bm r}-{\bm z})^2}}\\ &{}& \quad
\Big\{-T(Y,r)+T(Y,z)+T(Y,|{\bm r}-{\bm z}|)-T(Y,z)T(Y,|{\bm r}-{\bm
 z}|)\Big\}.\nonumber \eeq
Here $\abar\equiv \alpha_sN_c/\pi$, and ${\bm z}$ and ${\bm r}-{\bm z}$
are the transverse sizes of the two dipoles into which the parent dipole
${\bm r}$ has dissociated before scattering off the target. The last
term, quadratic in $T$, in the r.h.s. of the equation describes multiple
scattering (the simultaneous scattering of both daughter dipoles) and is
responsible for unitarization.

With this last term omitted, \eqref{BK} reduces to the BFKL equation,
which describes the unlimited (exponential) growth of the
scattering amplitude with $Y$ and the symmetric expansion of the support
of $T(Y,r)$ in $r$ towards both small and large dipole sizes. Note
however that the transverse non--locality in \eqnum{BK} is quite weak ---
the large daughter dipoles with size $z\gg r$ are suppressed by the
`dipole kernel' ${r^2} / [z^2 ({\bm r}-{\bm z})^2]$, whereas the very
small ones, with $z\ll r$, are disfavoured by the $T$--dependent terms in
the r.h.s., which exactly cancel each other as $z\to 0$ --- and can be
described as {\em diffusion} in the logarithmic variable
$\rho\equiv\ln(r_0^2/r^2)$. Here, $r_0$ is an arbitrary scale of
reference (say, the unitarization scale in the target at $Y=0$).

However, the fully non--linear equation (\ref{BK}) preserves (and
actually saturates) the unitarity bound $T\le 1$, as it manifestly has
$T=1$ as a fixed point. Because of that, the respective evolution is {\em
asymmetric} in $r$ (or $\rho$), and the solution $T(Y,r)\equiv T(Y,\rho)$
looks like a {\em front}, which interpolates between $T=1$ at relatively
small $\rho$ and $T=0$ at $\rho\to\infty$, and which with increasing $Y$
propagates towards larger values of $\rho$. Behind the front, the
scattering amplitude has reached the `black disk' limit $T=1$ and thus
cannot grow anymore. Ahead of the front, the amplitude is still weak,
$T\ll 1$, so the non--linear term in Eq.~(\ref{BK}) is unimportant and
the amplitude can grow according to the linear, BFKL, evolution. The
position $\rho_s(Y)\equiv \ln (r_0^2 Q_s^2(Y))$ of the front at `time'
$Y$, i.e. the value of $\rho$ where $T$ becomes of $\order{1}$,
represents the scale where unitarity corrections become important at
rapidity $Y$ and defines the {\em saturation momentum} $Q_s^2(Y)$.

The previous discussion already suggests that the progression of the
saturation front towards larger values of $\rho$ is driven by the BFKL
evolution of the dilute tail at $\rho \gg \rho_s(Y)$ --- the front is
`pulled' by its tail. This is an important property, as it allows us to
determine the position $\rho_s(Y)$ of the front and its shape around
$\rho_s(Y)$ from studies of the linear, BFKL, evolution alone. This
property is highly non--trivial, in view of the non--locality of the
non--linear equation (\ref{BK}). Other non--linear equations which are
non--local and exhibit saturation are known to develop a {\em pushed
front}, i.e., a front whose progression is driven by the growth and
accumulation of `matter' behind the front \citep{Saar}.

For the case of the BK equation with {\em fixed coupling} and for {\em
sufficiently high energy}, the pulled--front property follows from the
identification, made in Ref.~\citep{Munier:2003vc}, between the asymptotic
form of the BK equation at high energy\footnote{This is obtained via the
gradient expansion of the non--locality in Eq.~(\ref{BK}) to second order
in $\del/\del\rho$ (`diffusion approximation').} and the FKPP equation
(from Fisher Kolmogorov, Petrovsky, and Piscounov) of statistical
physics, for which this property has been established with mathematical
rigor\footnote{The FKPP equation describes a `reaction--diffusion process' in
the mean field approximation corresponding to very large occupation
numbers at saturation; see e.g. the review paper \citep{Saar}.}. However, this
identification does not extend to a {\em running} coupling, and hence it
fails to apply for the real QCD problem.

The running of the coupling is known to have dramatic consequences for
the high--energy evolution
\citep{Iancu:2002tr,Mueller:2002zm,Triantafyllopoulos:2002nz,Dumitru:2007ew},
and in particular for the approach towards saturation: the growth of the
coupling with decreasing momenta, or increasing dipole sizes, amplifies
the contribution of the latter to the evolution, which then becomes
asymmetric even in the absence of saturation. In fact, the BFKL
evolution with running coupling is infrared--unstable, in the sense that
it requires an infrared cutoff to avoid the blow--up of the QCD coupling
at $k_\perp\sim \Lam$, and then the results of the evolution are strongly
sensitive to the value of this cutoff --- so that the whole procedure has
no predictive power (see Fig.~\ref{fig:bfklrunres2} below). In that
scenario (BFKL with running coupling), the growth of the gluon
distribution at high momenta $k_\perp\gg \Lam$ is mostly feeded by
radiation from the bulk of the distribution at infrared ($k_\perp\sim
\Lam$) momenta. Hence, the whole perturbative framework becomes
questionable and, besides, one may expect the associated saturation front
--- as generated after enforcing unitarity --- to be of the `pushed'
type.

Yet, as our explicit numerical solutions will demonstrate, this is
actually not the case: the saturation front remains of the `pulled' type
even for a running coupling. This is so because the gluon modes with
$k_\perp\lesssim Q_s(Y)$ become inert due to saturation, so the evolution
is again driven by the dilute tail at high momenta, so like for fixed
coupling. In particular, the infrared problem is automatically avoided:
the saturation scale effectively acts as an infrared cutoff, which
becomes `hard' ($Q_s^2(Y)\gg\Lam^2$) for sufficiently high energy. This
opens the way towards realistic studies of the front dynamics within the
context of the {\em linear} evolution, as originally suggested in
Refs.~\citep{Iancu:2002tr,Mueller:2002zm}
. To that aim, the linear
evolution equations must be supplemented with an appropriate {\em
saturation boundary condition}, that we now describe.

Such a boundary condition must ensure that the amplitude never
becomes bigger than one. By itself, the position of the front is not {\em
a priori} known, but must be determined when solving the equation. To
that aim, let us first introduce a line of constant amplitude
$\rho=\rho_c(Y)$ via the condition
\beq\label{cline}
  T(Y,\rho=\rho_c(Y))\,=\,c\,,
 \eeq
where the number $c$ is strictly smaller than one, but not {\em much}
smaller. (The {\em saturation line} $\rho_s(Y)$ would correspond to
$c\sim 1$.) For $\rho <\rho_c(Y)$ and sufficiently high energy, the
solution $T_{\rm BFKL}(Y,\rho)$ to the BFKL equation would become larger
than one --- in fact, arbitrarily large. If this equation is to be solved
numerically, one may think about identifying the point $\rho_c(Y)$
numerically at each step in $Y$, and then enforcing the unitarity
limit $T=1$ for any $\rho$ which is smaller than $\rho_c(Y)$ and
sufficiently far away from it --- say, for $\rho\le \rho_c(Y)-\Delta$
with $\Delta\simeq \ln(1/c)$ a number of $\order{1}$. However, this would
not be a very good strategy in practice, since $T=1$ is not a fixed point
for the BFKL equation, so an amplitude of $\order{1}$ would be
exponentially amplified by the subsequent evolution. Even if, at small
$\rho$, one cuts off this evolution by hand step--by--step, it is not
clear (especially for running coupling) whether the spurious
radiation from small $\rho$ will not affect the tail of the front at
large $\rho$. It is therefore preferable, as originally suggested in
Ref.~\citep{Mueller:2002zm}
, to enforce the amplitude to {\em vanish} for
$\rho\le\rho_c(Y)-\Delta$ :
 \beq\label{BC}
  T(Y,\rho)\,=\,0\qquad\mbox{for}\qquad\rho\,\leq \,\rho_c(Y)-\Delta\,.
 \eeq
$T=0$ {\em is} a fixed point for the BFKL equation, so no further
evolution is possible in the `saturated domain' on the left of
$\rho_c(Y)-\Delta$, as it should. When decreasing $\rho$ below
$\rho_c(Y)$, the solution $ T(Y,\rho)$ will typically start by rising,
then reach a maximum $T_{\rm max}\sim\order{1}$, and eventually decrease
to zero. We shall conventionally identify the {\em saturation scale}
$\rho_s(Y)$ with the position of this maximum. In this procedure, the
numbers $c$ and $\Delta$ are to be viewed as free parameters, which are
correlated with each other, since $\Delta\sim \ln(1/c)$.

Since it yields, by construction, $T=0$ beyond the saturation front, this
procedure cannot be used for any physical problem which is sensitive to
the black disk limit, like deep inelastic scattering at low
$Q^2\lesssim Q_s^2(Y)$, or particle production at low transverse momenta.
On the other hand, as we shall see, this procedure accurately describes
the dynamics of the front, meaning its position and shape for $\rho
>\rho_c(Y)$, and hence it correctly provides the tail of the gluon
distribution at transverse momenta $k_\perp\ge Q_c(Y)$, with
$\rho_c(Y)\equiv \ln (r_0^2 Q_c^2(Y))$. Since $Q_c(Y)\simeq Q_s(Y)$, we
see that our results cover the phenomenologically interesting region
where geometric scaling is expected at high $Q^2$ (cf. Introduction).

To describe our numerical results, let us first change from the
coordinate to the momentum representation, i.e. from the scattering
amplitude $T(Y,r)$ to the `unintegrated gluon distribution'
$\mathcal{A}(Y,k)$ --- the quantity which enters the calculation of
cross--sections within the $k_T$--factorization. For the present
purposes, $\mathcal{A}(Y,k)$ can be defined as the following Fourier
transform of the dipole amplitude \citep{Iancu:2002tr}
 \beq\label{phiT} \mathcal{A}(Y,k)\,=\,
    \int \frac{\rmd^2\bm{r}}{2\pi r^2}\,
   {\rm e}^{-i \bm{k} \cdot \bm{r}}\,
    T(Y,r)\,.
    \eeq
With this definition, the more standard, `integrated', gluon distribution
is obtained as
 \beq\label{GDF}
x g(x,Q^2)\,=\,\frac{4N_c^2}{\pi^2\abar} \int^{Q^2}\!
\frac{\rmd^2\bm{k}}{(2\pi)^2}\int
 \rmd^2\bm{b}\ \mathcal{A}(Y,\bm{k},\bm{b})\,.\eeq
For our homogeneous target, $\mathcal{A}(Y,\bm{k},\bm{b})\equiv
\mathcal{A}(Y,k)$, hence the above integral over $\bm{b}$ simply yields the hadron
transverse area $\pi R^2$. With these conventions, the {\em gluon
occupation number} --- i.e., the number of gluons of a given color per
unit rapidity per unit volume in transverse phase--space --- is not
exactly $\mathcal{A}(Y,k)$, but rather $\mathcal{A}(Y,k)/\abar$ (up to a
numerical factor).

Via \eqnum{phiT}, the saturation front for $T(Y,r)$ translates into a
corresponding front for $\mathcal{A}(Y,k)$. For very large momenta,
$k\ggg Q_s(Y)$, one can use $T\sim r^2\ln(1/r^2)$ (`color transparency')
and then \eqnum{phiT} reproduces the bremsstrahlung spectrum:
$\mathcal{A}(Y,k)\sim 1/k^2$, or\footnote{From now on, we shall use the
notation $\rho$ for either $\ln(r_0^2/r^2)$, or $\ln(k^2/k_0^2)$ (with
$k_0=1/r_0$), the difference being clear from the context.}
$\mathcal{A}(Y,\rho)\sim \rme^{-\rho}$. This behaviour is modified when
$k$ gets closer to (but still larger than) $Q_s$, due to the BFKL
evolution in the presence of saturation. For instance, in the fixed
coupling case and for sufficiently high energy ($\abar Y\gg 1$), one
finds \citep{Iancu:2002tr,Mueller:2002zm,Munier:2003vc}
 \beq\label{Arho} \mathcal{A}(Y,\rho) \, \sim \,(\rho-\rho_s)\,
 {\rm e}^{-\gamma_s (\rho-\rho_s)}\,
 \exp\left\{-\frac{(\rho-\rho_s)^2} {2\beta \bar\alpha Y}
 \right\}\quad\mbox{for}\quad 1< \rho-\rho_s\lesssim 2\rho_s\,.\eeq
The exponential $\sim \rme^{-\gamma_s\rho}$ describes the modification of
the bremsstrahlung spectrum due to the BFKL `anomalous dimension'
$1-\gamma_s$, while the last, Gaussian, factor describes BFKL diffusion.
The numbers $\gamma_s\approx 0.63$ and $\beta\approx 48.52$ are specific
for the problem at hand: they characterize the BFKL evolution along a
line of constant amplitude, cf. \eqnum{cline} (see \citep{Iancu:2002tr,Mueller:2002zm}
for details). Also, the presence of
the saturation scale $\rho_s$ and the overall factor $\rho-\rho_s$ in the
r.h.s. of \eqnum{Arho} are the hallmarks of saturation, and can be
generated within the framework of the linear evolution only after
imposing the saturation boundary condition (\ref{BC}) \citep{Mueller:2002zm}
. Under the same conditions, the saturation scale is
obtained as
 \beq
 \rho_s(Y)\,=\, \lambda\abar Y\, -
 \,\frac{3}{2\gamma_s}\,\mathrm{ln}Y
 \,+\,\mbox{const.}
 \label{satmom}
 \eeq
where $\lambda\approx 4.88$
and the last, constant, term is not under control (since \eqnum{satmom}
is merely an asymptotic expansion at high--energy).

In the opposite limit of very low momenta $k\ll Q_s(Y)$, the integral in
\eqnum{phiT} is dominated by large dipole sizes $r\gg 1/Q_s$ for which
$T=1$; one then finds $\mathcal{A}(Y,k)\simeq \ln[Q_s(Y)/k] =
[\rho_s(Y)-\rho]/2$. Thus, behind the front, $\mathcal{A}(Y,k)$ is not
exactly constant (unlike the dipole amplitude), but it is slowly growing
--- logarithmically in both $1/k$ and $Y$. What saturates at high density
is not the gluon occupation number $\mathcal{A}/\abar$, but rather the
rate for gluon emission \citep{CGCreviews}.


The analytic results in Eqs.~(\ref{Arho}) and (\ref{satmom}), together
with the corresponding ones at running coupling
\citep{Mueller:2002zm,Triantafyllopoulos:2002nz}
, have been already tested
in the literature against numerical solutions to the BK equation. Here,
we are rather interested to compare the solutions to the latter against
numerical solutions for the BFKL equation supplemented with the boundary
condition (\ref{BC}) --- with both fixed and running coupling. The
momentum--space version of the BK equation is particularly simple in that
the non--linear term is local:
\beq
  \frac{\del}{\del Y}\,\mathcal{A}(Y,k)=\bar\alpha_s
\int {\rmd^2\bm{q}\over \pi}\,
 {1 \over q^2 (\bm{k}-\bm{q})^2}\,
\left(q^2\mathcal{A}(Y,q) -\,{k^2\over 2}\,\mathcal{A}(Y,k)\right)
-{\bar\alpha_s}\big(\mathcal{A}(Y,k)\big)^2 \,.
 \label{BKA} \eeq
The linear version of this equation, i.e., the BFKL equation for
$\mathcal{A}(Y,k)$, will be solved with an absorptive boundary condition
similar to that for $T(Y,r)$ in \eqnum{BC}. This is appropriate since,
with the normalization in \eqnum{phiT}, the saturation effects in the
gluon distribution become important when $\mathcal{A}(Y,k)\sim\order{1}$.

For the fixed coupling calculations we shall use $\abar=0.2$. To include
running coupling effects, we shall pull the $\abar$ factor inside the
$q$--integral in \eqnum{BKA} and use the one--loop expression for the
running coupling with scale $Q^2={\rm max}(k^2, q^2)$ and $\Lam=
200$~MeV. This simple prescription is in agreement with the recently
constructed running--coupling version of the BK equation \citep{Balitsky:2006wa,Kovchegov:2006vj}
. To avoid the infrared divergence
of the coupling at $Q^2=\Lam^2$, we shall replace $\alpha_s(Q^2) \to
\alpha_s(Q^2+\mu^2)$ for some parameter $\mu$. Our default choice will be
$\mu^2 = 0.5$ GeV$^2$, but we shall study the sensitivity of our results
to variations in $\mu$. Our initial condition $\mathcal{A}(Y=0,k)$ is
given by the bremsstrahlung spectrum for $k > 1$~GeV (with maximal height
$\mathcal{A}=0.5$) and it vanishes for $k < 1$~GeV.

\begin{figure}[t] {\centerline{
  \includegraphics[angle=270, scale=0.6]{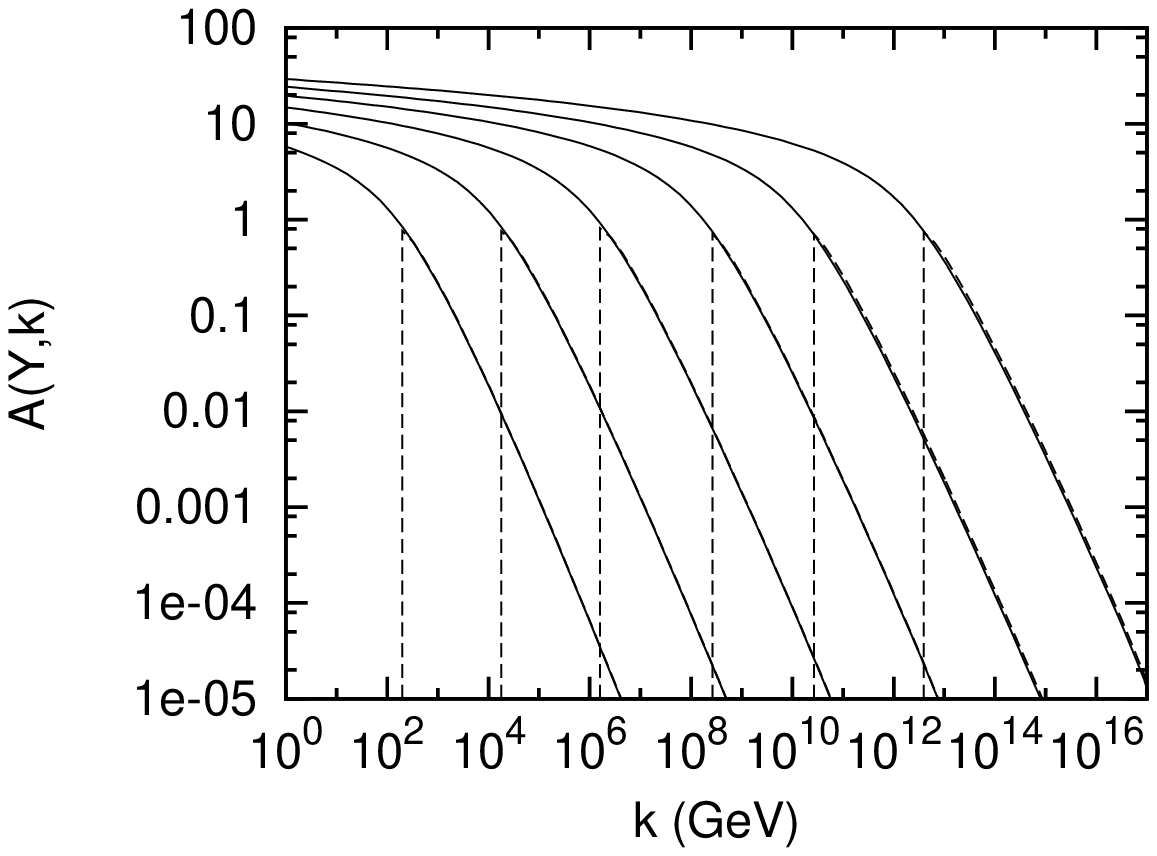}
  \includegraphics[angle=270, scale=0.6]{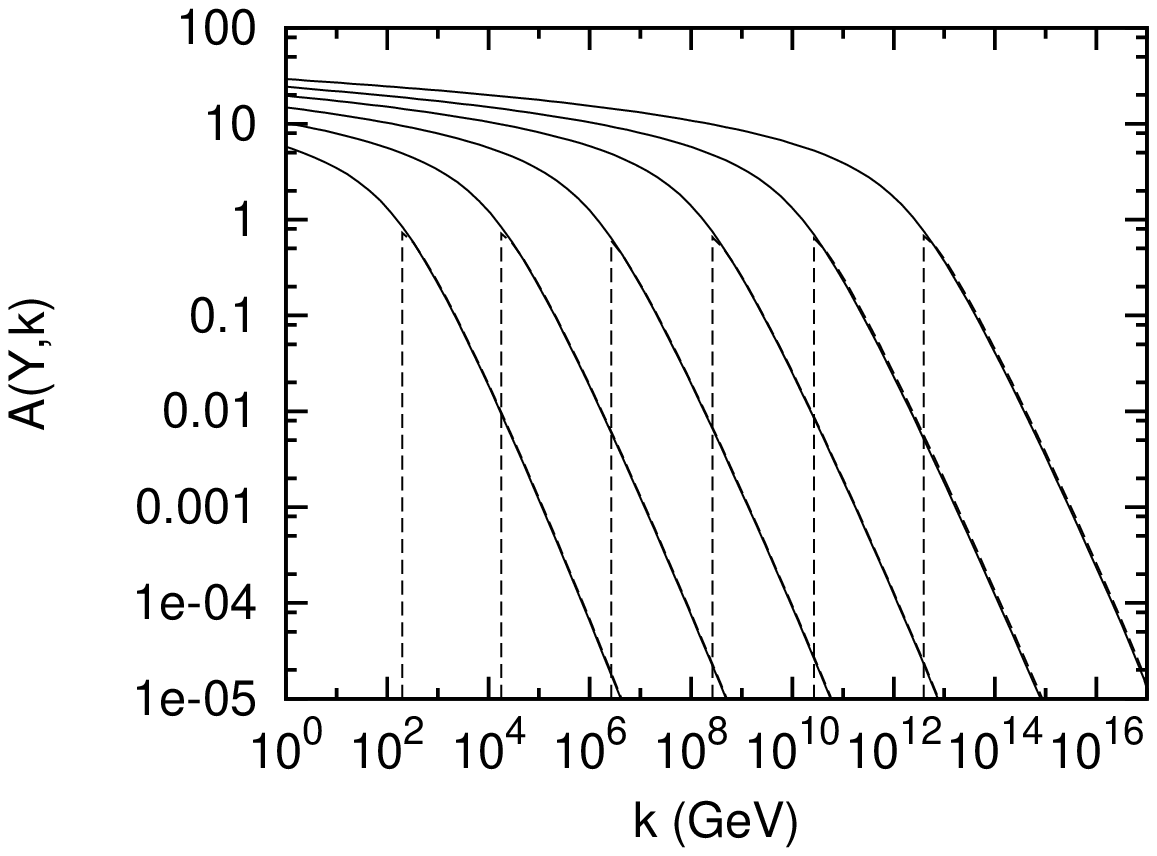}}
\caption {\label{fig:bfklres1} \sl The solid lines are the solutions of the
BK equation (\ref{BKA}) while the dashed lines have been obtained by
applying the absorptive boundary (\ref{BC}) on the BFKL equation. The
leftmost curve has $Y=20$ and $Y$ is increased by 10 units for each new
curve. For the figure on the left we have chosen $c=0.1$ and
$\Delta=5.0$, while for the figure on the right we have $c=0.3$ and
$\Delta = 3.0$. }
 \label{fig:Afix}}
\end{figure}

For the fixed coupling case, our results are displayed in
Fig.~\ref{fig:Afix} for five values of the rapidity within the range
$20\le Y\le 70$ and for two different choices for the parameters $c$ and
$\Delta$. The most important observation about these results is that the
saturation fronts generated by the two types of evolution do precisely
coincide with each other for all momenta $\rho\ge \rho_c(Y)-\Delta$, and
for all the considered rapidities. This property is not altered by
changing the values for $c$ and (correlated to it) for $\Delta$. We have
also checked that these numerical curves are consistent with the analytic
estimates in Eqs.~(\ref{Arho}) and (\ref{satmom}), including the expected
values for $\lambda$ and $\gamma_s$.

\begin{figure}[t] {\centerline{
  \includegraphics[angle=270, scale=0.6]{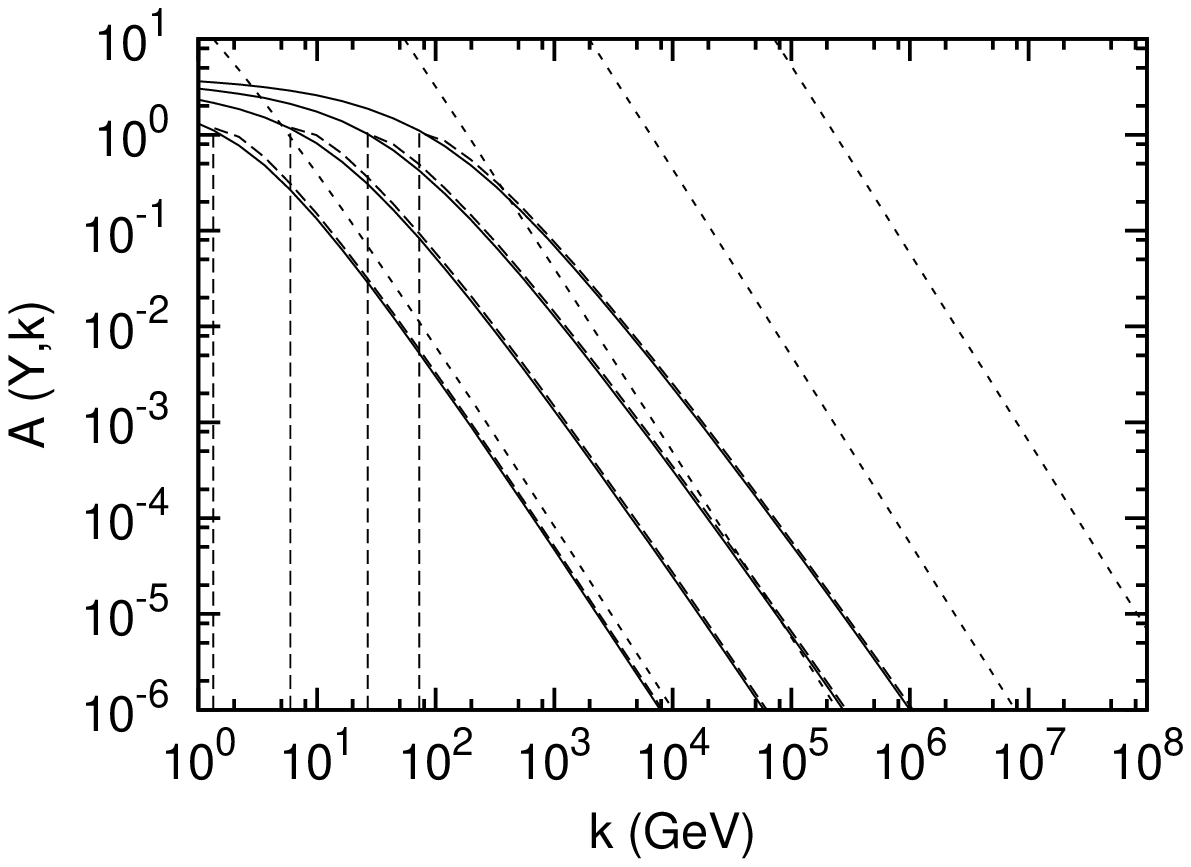}
  \includegraphics[angle=270, scale=0.6]{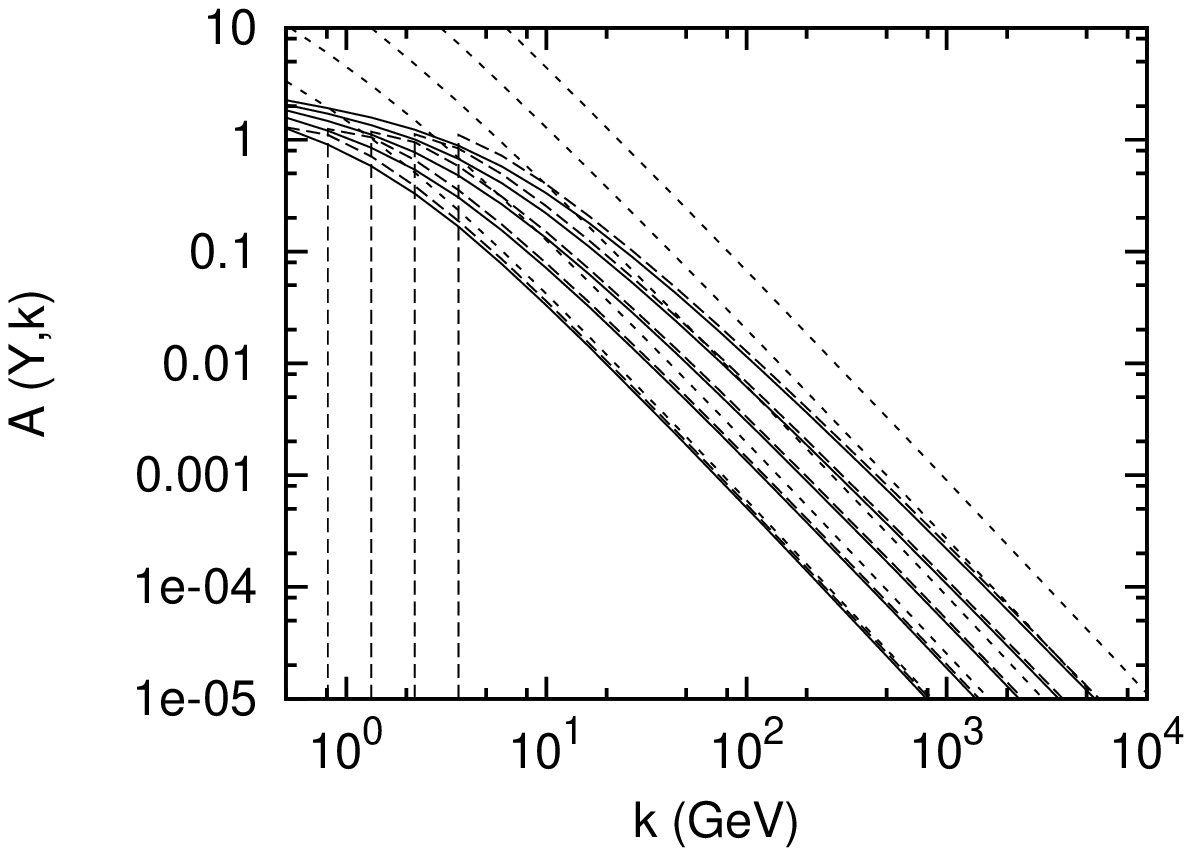}}
\caption {\sl
The running coupling results for: BK (solid curves), BFKL
with absorptive boundary (long dashed curves) and pure BFKL (short dashed
curves) for (left) $Y= 10, 20, 30$ and $40$, and (right) $Y=6, 8, 10, 12$ and $14$.
 For the absorptive boundary we used $c=0.1$ and $\Delta=5.0$.
\label{fig:bfklrunres1} }}
\end{figure}

We now turn to the more realistic case of a running coupling. Then, as
alluded to before, the pure BFKL evolution is infrared unstable, since
the rapid growth of the gluon distribution at small values of $k$ (where
is coupling is larger) is also feeding the growth at higher $k$.
Therefore the linear evolution behaves quite differently compared to the
non-linear one, even at high $k$. This is clearly visible in
Fig.~\ref{fig:bfklrunres1} where we compare the strict BFKL evolution to
the BK one, and to the BFKL evolution with the absorptive boundary. Once
again, there is a perfect matching between the saturation fronts provided
by BK and, respectively, BFKL with saturation boundary. (For the latter,
we used the same values for $c$ and $\Delta$ as at fixed coupling.) On
the other hand, we see a dramatic difference with respect to the linear
evolution, which progresses much more rapidly towards the right.

From these curves, it is also possible to extract the $Y$--dependence of
the saturation momentum $\rho_s(Y)$ for running coupling. We find that
the squared--root law $\rho_s \simeq \lambda_r \sqrt{Y}$ predicted by the
theory \citep{Iancu:2002tr,Mueller:2002zm,Triantafyllopoulos:2002nz}
for
asymptotically high energies provides a good fit to our numerical results
for $Y\ge 10$, with a fitted value $\lambda_r\simeq 2.8$ which agrees
reasonably well with the (asymptotic) theoretical
expectation\footnote{Specifically, for asymptotically large $Y$, the
running--coupling BFKL evolution yields \citep{Iancu:2002tr,Mueller:2002zm}
: $\rho_s(Y)\simeq \sqrt{2\lambda
b_0Y}$ where $\lambda\simeq 4.88$ is the same number as in \eqnum{satmom}
and $b_0\equiv {12N_c}/({11N_c -2N_f})$ is the coefficient in the
one--loop running coupling: $\bar\alpha(Q^2)={b_0}/{\ln (Q^2/\Lam^2)}$.
In our simulations, we use $N_f=0$, hence we expect $\lambda_r\equiv
\sqrt{2\lambda b_0}\simeq 3.26$, which is indeed consistent with the fit
to the curves in Fig.~\ref{fig:bfklrunres1}.} $\lambda_r\simeq 3.2$.

Now, from the phenomenological point of view, we are more interested in
values of $Y$ which are not that large, say $Y\le 14$ (corresponding to
$x \gtrsim 10^{-6}$), as relevant for forward jet production at LHC. With
that in mind, we also show in Fig.~\ref{fig:bfklrunres1} (right) the
results for lower values of $Y$, between 6 and 14 units; one can thus see
that the absorptive boundary method works equally well also for such
lower rapidities.

Finally, to illustrate the infrared stability introduced by saturation,
we exhibit in Fig.~\ref{fig:bfklrunres2} results obtained for different
values of the IR cutoff $\mu^2$ inserted in the running coupling. Unlike
the pure BFKL results (left figure), which are extremely sensitive to a
change in $\mu$, the results corresponding to the saturation boundary
condition (right figure) show no sensitivity whatsoever.

\begin{figure}[t] {\centerline{
  \includegraphics[angle=270, scale=0.6]{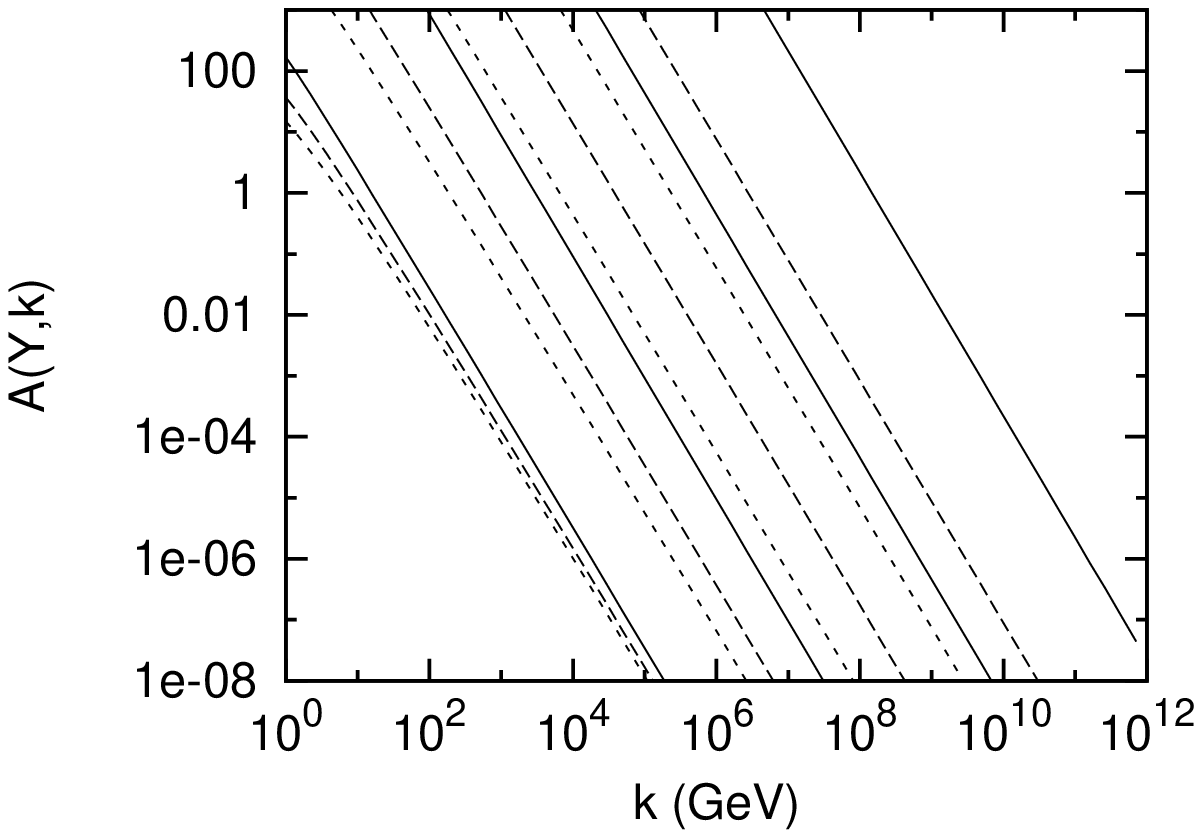}
  \includegraphics[angle=270, scale=0.6]{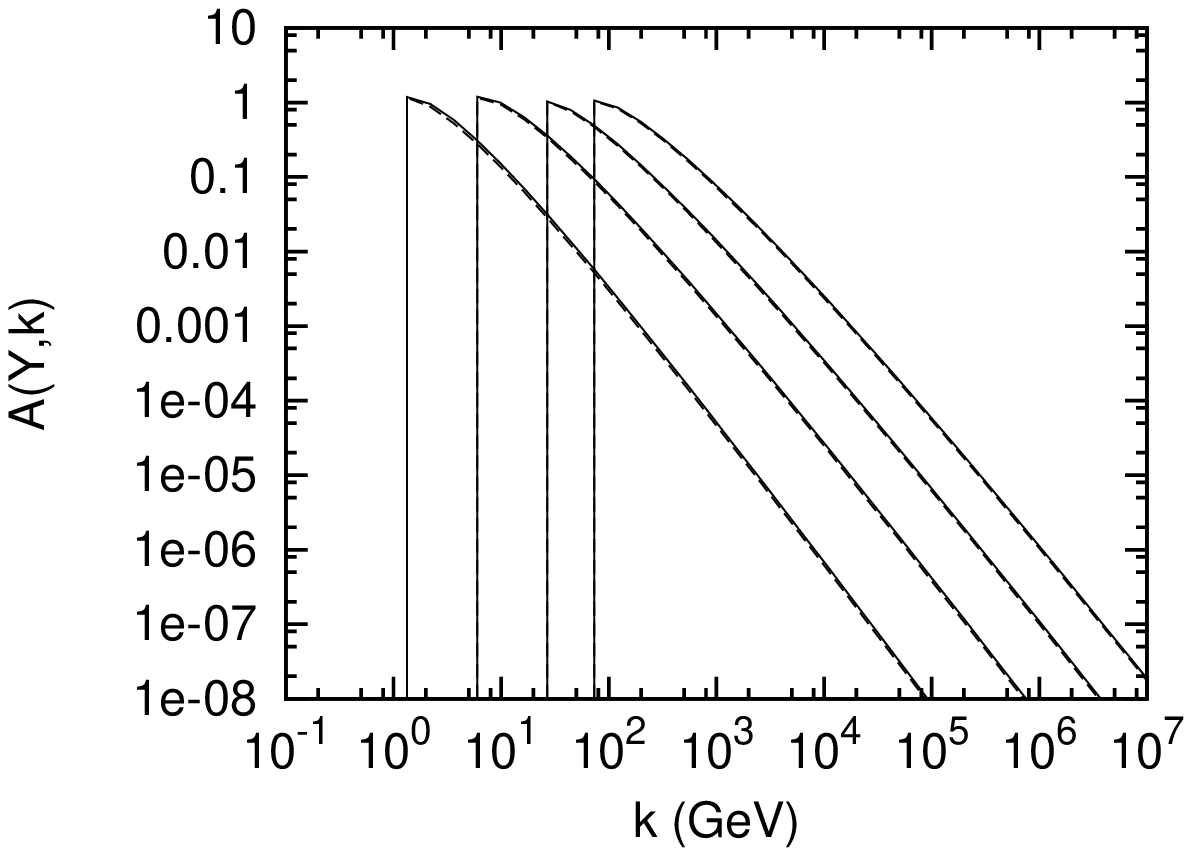}}
\caption {\sl Testing the sensitivity of the BFKL evolution to the IR cutoff
$\mu$. Left: The pure BFKL evolution with $\mu^2 = 0.5$ GeV$^2$ (solid
lines), $\mu^2 = 1.0$ GeV$^2$ (long dashed lines) and $\mu^2 = 2.0$
GeV$^2$ (short dashed lines), for $Y=10, \,20, \,30$ and 40. Right: The
BFKL evolution with absorptive boundary for $\mu^2 = 0.5$ GeV$^2$ (solid
lines), $\mu^2 = 2.0$ GeV$^2$ (dashed lines) and for the same values for $Y$
as before. \label{fig:bfklrunres2} }}
\end{figure}

\section{CCFM evolution with absorptive boundary}

In this section we shall present a compact version of the CCFM equation
\citep{CCFM} to which we shall apply the boundary condition described in
the previous section. A more comprehensive discussion of the CCFM
formalism and its relation to BFKL will be given elsewhere \citep{Avsar},
together with more detailed numerical studies, of which the present
Letter is only giving a glimpse.

As mentioned in the introduction, the CCFM evolution takes into account
the quantum coherence between successive emissions via angular ordering
in the parton cascades. Accordingly, the respective gluon distribution
now depends on three variables, $\mcal{A} = \mcal{A}(x, k, \bar{q})$,
where the third variable $\bar{q}$ is a transverse momentum related to
the maximum angle which determines the phase space where emissions are
allowed. This angle is set by the hard scattering of the space--like
photon against a quark inside the proton. It is customary to define the
variable $\xi$ which is the squared angle, $\xi \equiv q^2/(y^2E^2)$,
where $q$ is the transverse momentum of a gluon emitted in the
$s$--channel, $y$ is its longitudinal momentum fraction, and $E$ is the
energy of the proton; then, all emissions must satisfy  $\xi \leq
\bar{\xi} \equiv \bar{q}^2/(x^2E^2)$.

The CCFM equation for $\mcal{A}$ can be written in different versions,
depending on how `exclusive' we choose the gluon distribution to
be\footnote{Recall that $\mcal{A}(x, k)$ represents the distribution of
gluons produced by the evolution in the $t$--channel, as available for
the interaction with an external projectile.}. That is, so long as one is
not interested in the structure of the final state, one can `integrate
out' some of the emissions in the $s$--channel, as they do not change the
overall (unintegrated) gluon distribution (and hence neither the
probability for the interaction with a projectile). In practice, this
amounts to suitable cancelations between `real' gluon emissions and
`virtual' terms (or `Sudakov factors'). Of course, if one is interested
in studying the exclusive final states, all emissions which were removed
from the initial state must later be included as final state radiation.

\comment{ If one keeps within $\mcal{A}$ only the emissions associated
with the $1/z$ pole in the splitting function, then the CCFM equation can
be written as the following integral equation
  \begin{eqnarray}
\mathcal{A}(x, k, \bar{p}) = \abar \int_x^1 \frac{\rmd z}{z}
\int \frac{\rmd^2\bm{p}}{\pi p^2} \,\theta (\bar{p} - zp)\Delta_{ns}(k,z,p)
\mathcal{A}\left(\frac{x}{z}, |\bm{k}+(1-z)\bm{p}|, p\right)\,,
\label{eq:ccfminteq2}
 \end{eqnarray}
where the variable $\bm{p}$ is in fact the `rescaled momentum',
$\bm{p}\equiv \bm{q}/(1-z)$, where $\bm{q}$ is the actual transverse
momentum and $z$ the energy fraction taken by the virtual gluon emitted
in the $t$--channel; in particular, $\bar{p}=\bar{q}/(1-x)$. The theta
function comes from the angular ordering constraint $\xi \leq \bar{\xi}$.
There is also an energy ordering implicit in \eqref{eq:ccfminteq2}:
emissions are ordered in energy as well as in angle. In what follows,
however, we are only interested in the small--$x$ behaviour, so we make
no distinction between $p$ and $q$ and assume the energy ordering to be
automatic. Finally, $\Delta_{ns}$ is the so--called `non--Sudakov form
factor', which accounts for virtual corrections and is necessary to
ensure probability conservation }

 If one keeps within $\mcal{A}$ only the
emissions associated with the $1/z$ pole in the splitting function, then
the CCFM equation can be written as the following integral equation
  \begin{eqnarray}
\mathcal{A}(x, k, \bar{q}) = \abar \int_x^1 \frac{\rmd z}{z}
\int \frac{\rmd^2\bm{q}}{\pi q^2} \,\theta (\bar{q} - zq)\Delta_{ns}(k,z,q)
\mathcal{A}\left(\frac{x}{z}, |\bm{k}+\bm{q}|, p\right)\,,
\label{eq:ccfminteq2}
 \end{eqnarray}
where $\bm{q}$ and $1-z$ are, respectively, the transverse momentum and
the energy fraction of the `real' gluon emitted (in one step of the
evolution) in the $s$--channel\footnote{Strictly speaking,
\eqnum{eq:ccfminteq2} should involve the rescaled variables
$\bar{p}=\bar{q}/(1-x)$ and $\bm{p}\equiv \bm{q}/(1-z)$, but here  we are
only interested in the small--$x$ behaviour, so we make no distinction
between e.g. $p$ and $q$ \citep{Avsar}
.}. The theta function comes from
the angular ordering constraint $\xi \leq \bar{\xi}$. There is also an
energy ordering implicit in \eqref{eq:ccfminteq2}: emissions are ordered
in energy as well as in angle. Finally, $\Delta_{ns}$ is the so--called
`non--Sudakov form factor', which accounts for virtual corrections and is
necessary to ensure probability conservation
\begin{eqnarray}
\Delta_{ns}(z,k,q) = \exp \left( -\abar \int_{z}^{1}\frac{\rmd z'}{z'}
\int_{z'^2q^2}^{k^2} \frac{\rmd q'^2}{q'^2} \right ) =
\exp \left( -\abar  \log \left(\frac{1}{z}\right)\log
\left(\frac{k^2}{zq^2}\right) \right ).
\label{eq:nonsud}
\end{eqnarray}

Since $\mcal{A}$ now depends upon three variables, \eqnum{eq:ccfminteq2}
is much more difficult to solve than the BFKL equation. To simplify the
numerics and thus be able to explore a relatively wide range in $k$ and
$Y$, it is convenient to use a simpler version of the CCFM equation, that
we shall now derive. This is possible since, as first noticed in
\citep{Andersson:1995ju} (see also Ref.~\citep{Salam:1999ft}), one can
further integrate out some of the real emissions explicit in
\eqnum{eq:ccfminteq2}, and thus get a more inclusive equation. The main
point, to be further detailed in \citep{Avsar}, is that the non--Sudakov
factor in \eqref{eq:nonsud}, despite its name, can be used as a kind of a
Sudakov factor, to cancel a certain class of real emissions\footnote{With
this we mean that the total probability to have any number of real
emissions of the given class times the form factor is one (the emissions
are probability conserving).}.

Since in CCFM (like in BFKL) we integrate over the transverse momentum
$q$ of the emitted gluon, one can roughly have three possibilities:
either $k' \approx k \gg q$, or $k' \approx q \gg k$, or, finally, $k
\approx q \gg k'$. (Here $k' = \vert \bm{k} + \bm{q} \vert$, so we
consider the elementary splitting $\bm{k}'\to \bm{k}+\bm{q}$.) The
emissions satisfying the first condition are the ones which can be
canceled against $\Delta_{ns}$. To that aim, we must also include, within
the integrand of \eqnum{eq:ccfminteq2}, the so--called \emph{kinematical
constraint} $k^2
> zq^2$ which ensures that the squared four--momenta of the virtual
propagators are dominated by their transverse part as required by the
multi--Regge kinematics. This constraint is generally kept implicit in
the CCFM (or BFKL) literature, since the Regge kinematics is guaranteed
{\em to the order of interest}\,; yet, its explicit inclusion in the
equations introduces corrections which are formally of higher order in
$\alpha_s$, but which can be numerically important. After including this
constraint, we can remove the factor $\Delta_{ns}$ from
\eqnum{eq:ccfminteq2} and simultaneously limit ourselves to emissions
satisfying the last two constraints written above, which can be
summarized as $\theta(q^2- \mathrm{min}(k^2, k'^2))$. We thus obtain
\begin{eqnarray}
\mathcal{A}(x, k, \bar{q}) = \abar \int_x^1 \frac{\rmd z}{z}
\int \frac{\rmd^2\bm{q}}{\pi q^2} \, \theta (\bar{q} - zq)\theta (k^2 - zq^2)
\theta(q^2- \mathrm{min}(k^2, k'^2))
\mathcal{A}\left(\frac{x}{z}, k', q\right).\nn
\end{eqnarray}
Since $\bar{q} \geq k$ for all cases of physical interest, we further
have $\bar{q}^2 \geq k^2 \geq zq^2 \geq z^2q^2$. Therefore the angular
ordering is automatic and $\theta (\bar{q} - zq)$ can be neglected. This
means that the dependence on the third variable $\bar{q}$ drops out, and
we can write (note also that $q \geq k'$)
\begin{eqnarray}
\mathcal{A}(x, k) = \abar \int_x^1 \frac{\rmd z}{z}
\int \frac{\rmd^2q}{\pi q^2} \, \theta (k^2 - zq^2)
\theta(q^2- \mathrm{min}(k^2, k'^2))
\mathcal{A}\left(\frac{x}{z}, k'\right)\,,
\end{eqnarray}
which as compared with the original \eqnum{eq:ccfminteq2} represents a
considerable simplification.

We shall now perform the integration over the azimuthal angle $\phi$
between $\bm{q}$ and $\bm{k}$. To that aim, it is convenient to replace
$\theta(k^2-zq^2)$ by $\theta(k^2-zk'^2)$, which is allowed within the
current approximations\footnote{Indeed, we have either $k' \approx q \gg
k$, in which case the replacement is obviously correct, or $k\approx q
\gg k'$, in which case both the first and the second theta function can
be replaced by 1.}, and then switch the integration variables from
$\bm{q}$ to $\bm{k}'$ and, respectively, from $z$ to $x/z$ (which we
rename as $z$). This yields
\begin{eqnarray} \mathcal{A}(x, k) = \abar \int_x^1
\frac{\rmd z}{z} \int_0^\pi \frac{\rmd\phi}{\pi}\int \frac{\rmd k'^2}{\vert
\bm{k}'-\bm{k}\vert^2} \, \theta (z- xk'^2/k^2) \,\theta\big(\vert \bm{k}' -
\bm{k} \vert^2- \mathrm{min}(k^2, k'^2)\big)
\mathcal{A}(z, k'). \nonumber \\
\end{eqnarray}
Notice that this equation is infrared finite. After also performing the
$\phi$ integral, one finds
  \begin{eqnarray}
\mathcal{A}(x, k) = \abar \int_x^1 \frac{\rmd z}{z}
\int \frac{\rmd k'^2}{\vert k'^2-k^2\vert} \, \theta (z- xk'^2/k^2)\,
h(\kappa )\,
\mathcal{A}(z, k')\,,
\label{eq:ccfminteq3}
\end{eqnarray}
where $\kappa \equiv \mathrm{min}(k^2,k'^2)/\mathrm{max}(k^2,k'^2)$ and
\begin{eqnarray}
h(\kappa) \equiv  1 -\frac{2}{\pi}\arctan\left(\frac{1+\sqrt{\kappa}}{1-\sqrt{\kappa}}\sqrt{
\frac{2\sqrt{\kappa}-1}{2\sqrt{\kappa}+1}} \right ) \theta(\kappa-1/4).
\end{eqnarray}
Notice that $h(\kappa)\to 0$ as $\kappa\to 1$, so \eqnum{eq:ccfminteq3} has indeed
no singularity at $k'=k$. To obtain an integro-differential
equation, we define $z=e^{-y}$
and $x=e^{-Y}$ and differentiate the l.h.s. with respect to $Y$. We thus
get
  \begin{eqnarray}
  \partial_Y\mathcal{A}(Y, k)& =& \abar \int
\frac{\rmd k'^2}{\vert k^2-k'^2\vert} h (\kappa )
\biggl( \theta(k^2-k'^2)\mathcal{A}(Y,k')  \nonumber \\
 &{}&\qquad +   \theta(k'^2-k^2)\theta(Y-\log(k'^2/k^2))
\mathcal{A}\big(Y- \log(k'^2/k^2),k'\big) \biggr)\,,
\label{eq:ccfmdiffeq3}
\end{eqnarray}
which is our final version for the CCFM equation and is equivalent to an
equation originally proposed in Ref.~\citep{Andersson:1995ju}
(although
there this was not derived from  the CCFM equation \eqref{eq:ccfminteq2}
in the way we did). The fact that this equation is nonlocal in $Y$ should
not come as a surprise, since the CCFM evolution is not really an
evolution in $Y$, but rather in $\xi$.


\comment{
\begin{eqnarray} \int_0^\pi \frac{d\phi}{\pi \vert
\bm{k}-\bm{k}'\vert^2}\, \theta(\vert \bm{k} - \bm{k}' \vert^2-
\mathrm{min}(k^2, k'^2)) \,=\, \frac{1}{\vert k^2 - k'^2\vert}\, h\left (
\frac{\mathrm{min}(k^2,k'^2)}{\mathrm{max}(k^2,k'^2)} \right )
\end{eqnarray}
}

\begin{figure}[t] {\centerline{
  \includegraphics[angle=270, scale=0.6]{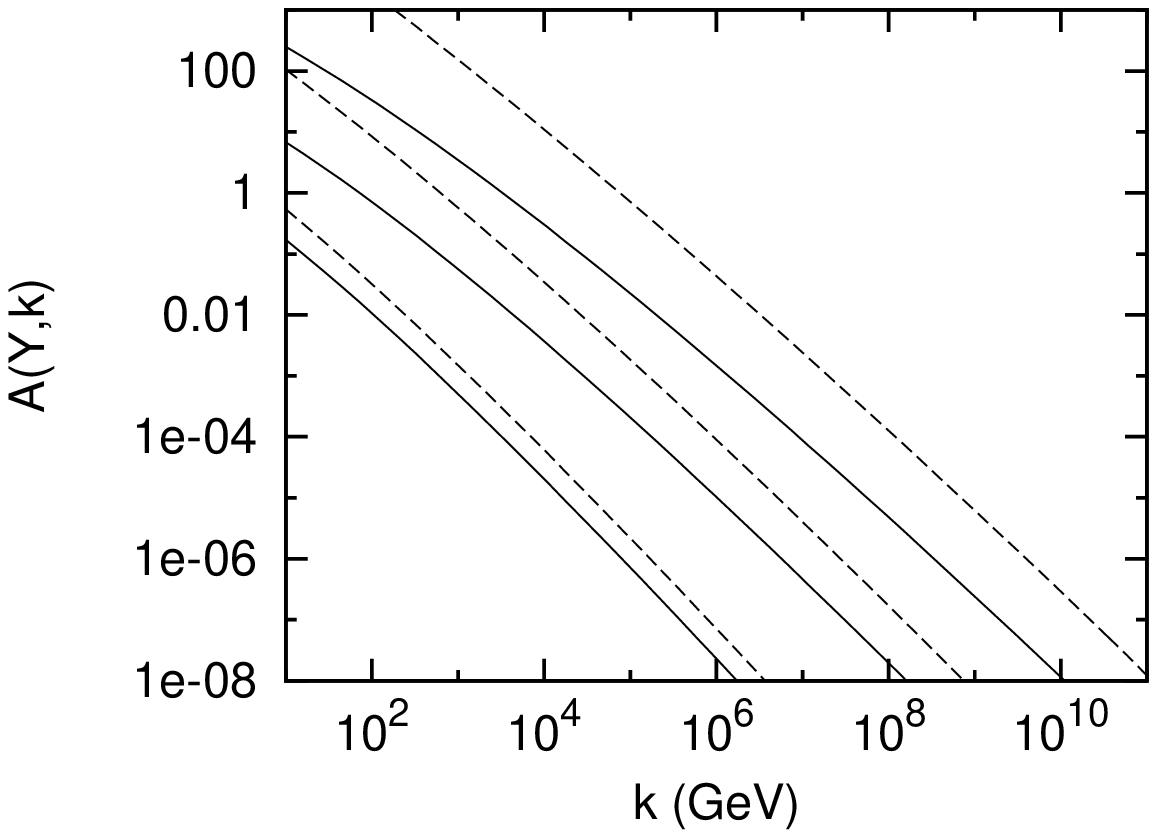}
  \includegraphics[angle=270, scale=0.6]{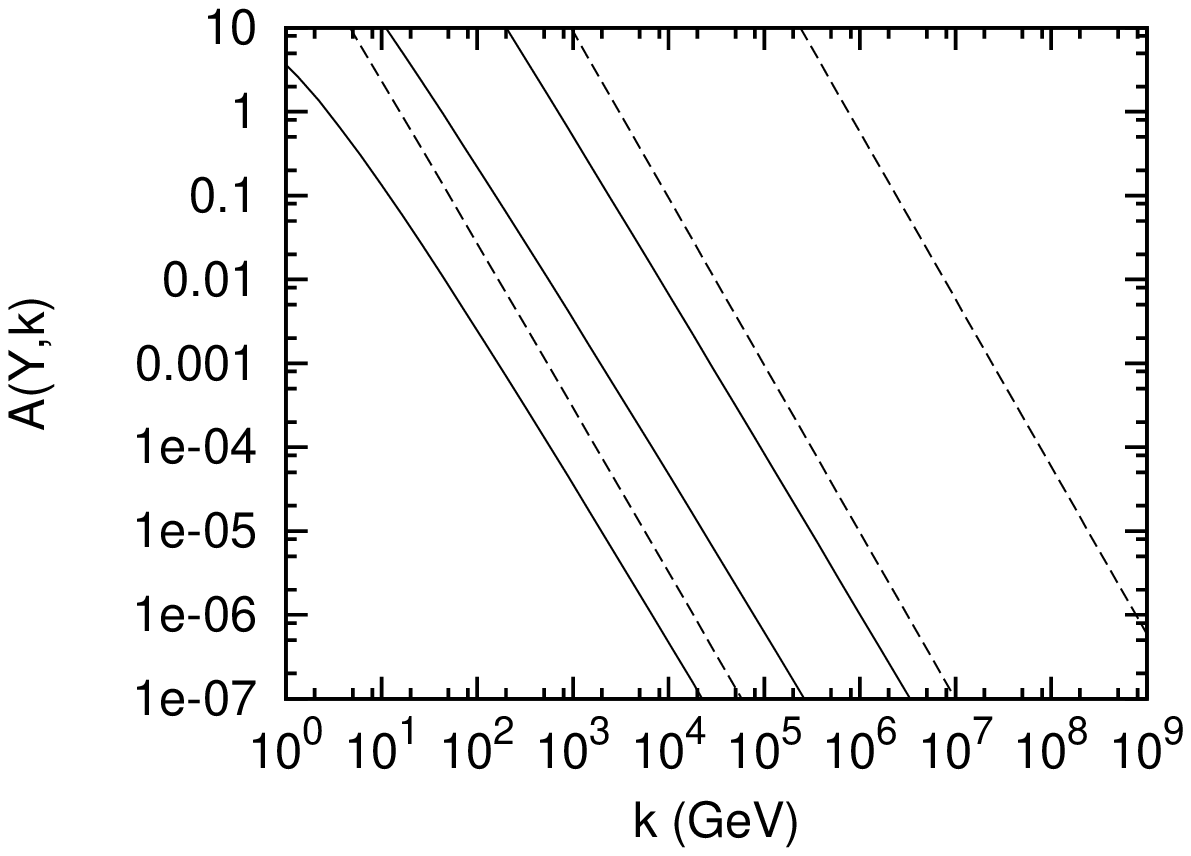}}
\caption {\sl Comparison between the solutions to the CCFM equation
\eqref{eq:ccfmdiffeq3} (solid lines) and the BFKL equation
 (dashed lines) for $Y=10,
\,20,$ and $30$. Left: fixed coupling $\abar=0.2$.
Right: running coupling.
  \label{fig:bfklres4} }}
\end{figure}

\begin{figure}[t] {\centerline{
  \includegraphics[angle=270, scale=0.6]{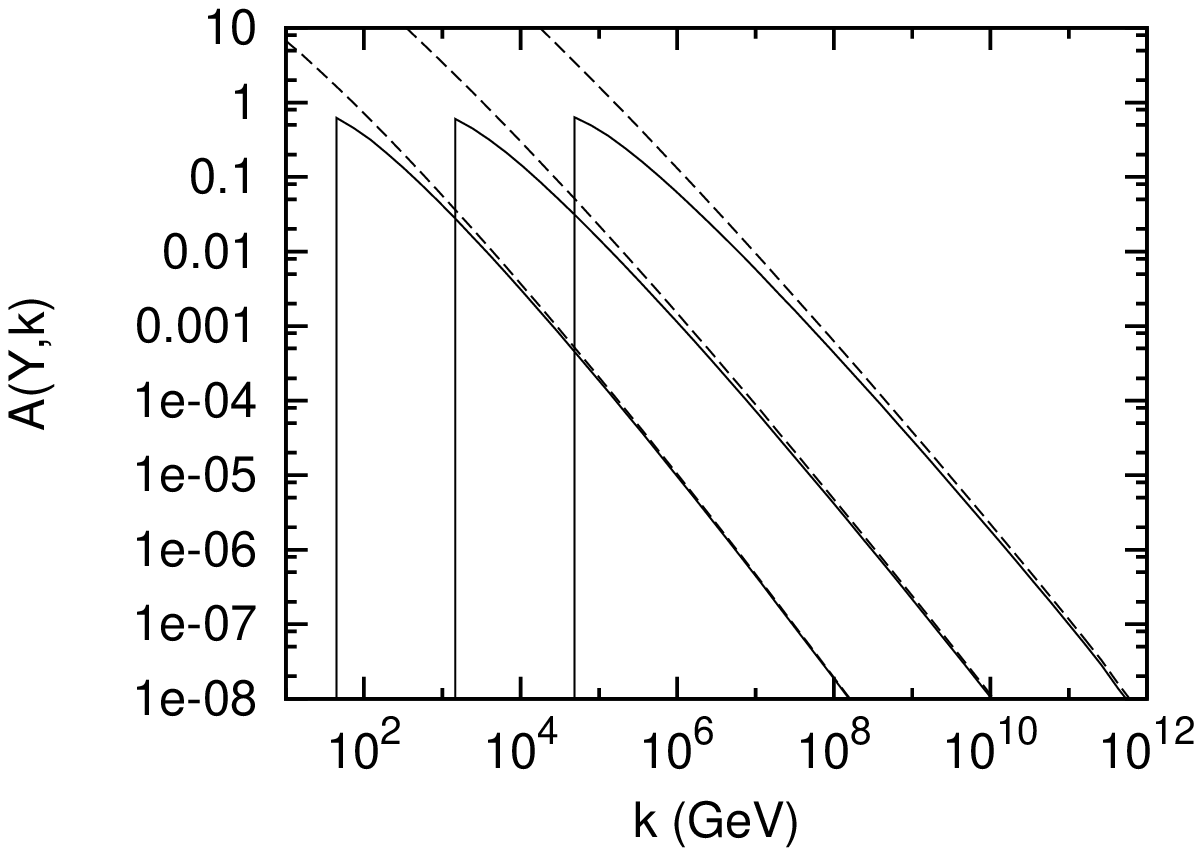}
  \includegraphics[angle=270, scale=0.6]{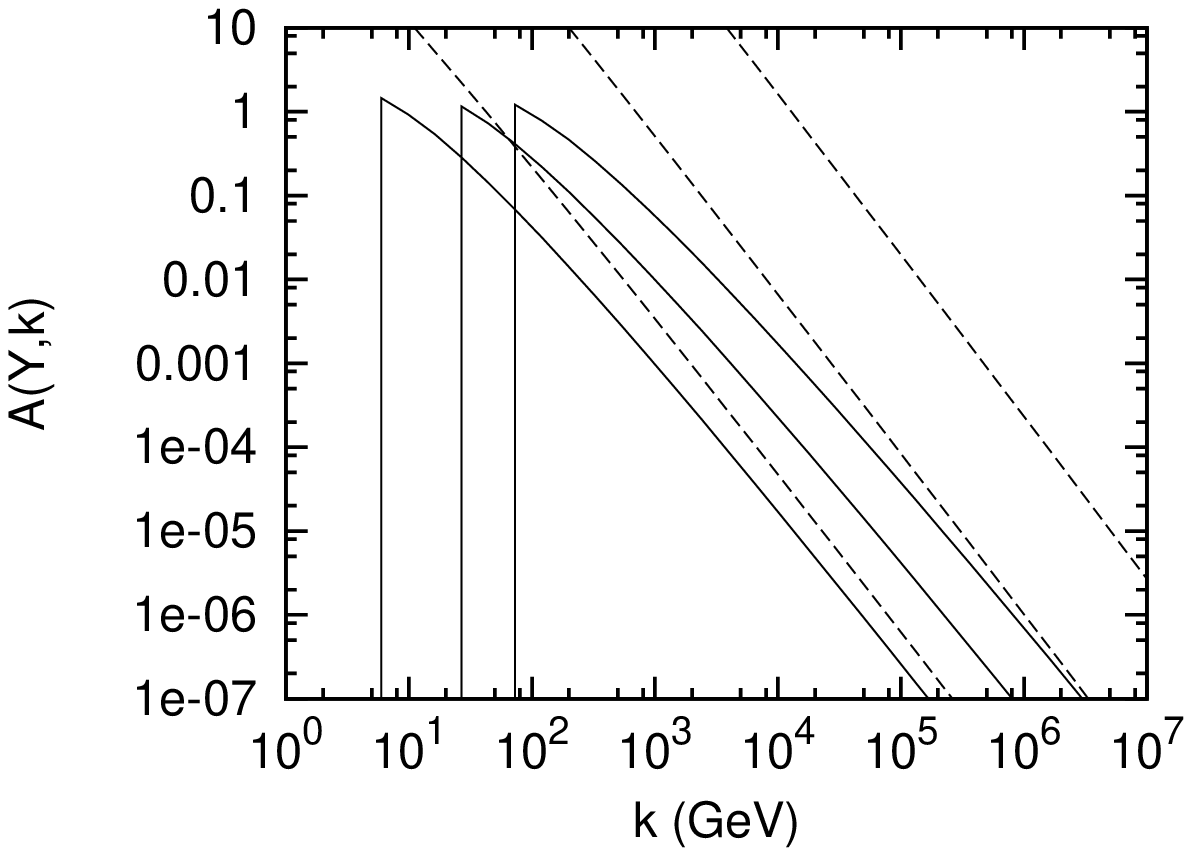}}
\caption {\sl Comparison between the solutions to the CCFM equation
\eqref{eq:ccfmdiffeq3} with saturation boundary (solid lines) and without
it (dashed lines), for $Y= 20, \,30,$ and $40$.
Left: fixed coupling $\abar=0.2$. Right: running coupling.
  \label{fig:bfklres5} }}
  \end{figure}

\begin{figure}[t] {\centerline{
  \includegraphics[angle=270, scale=0.6]{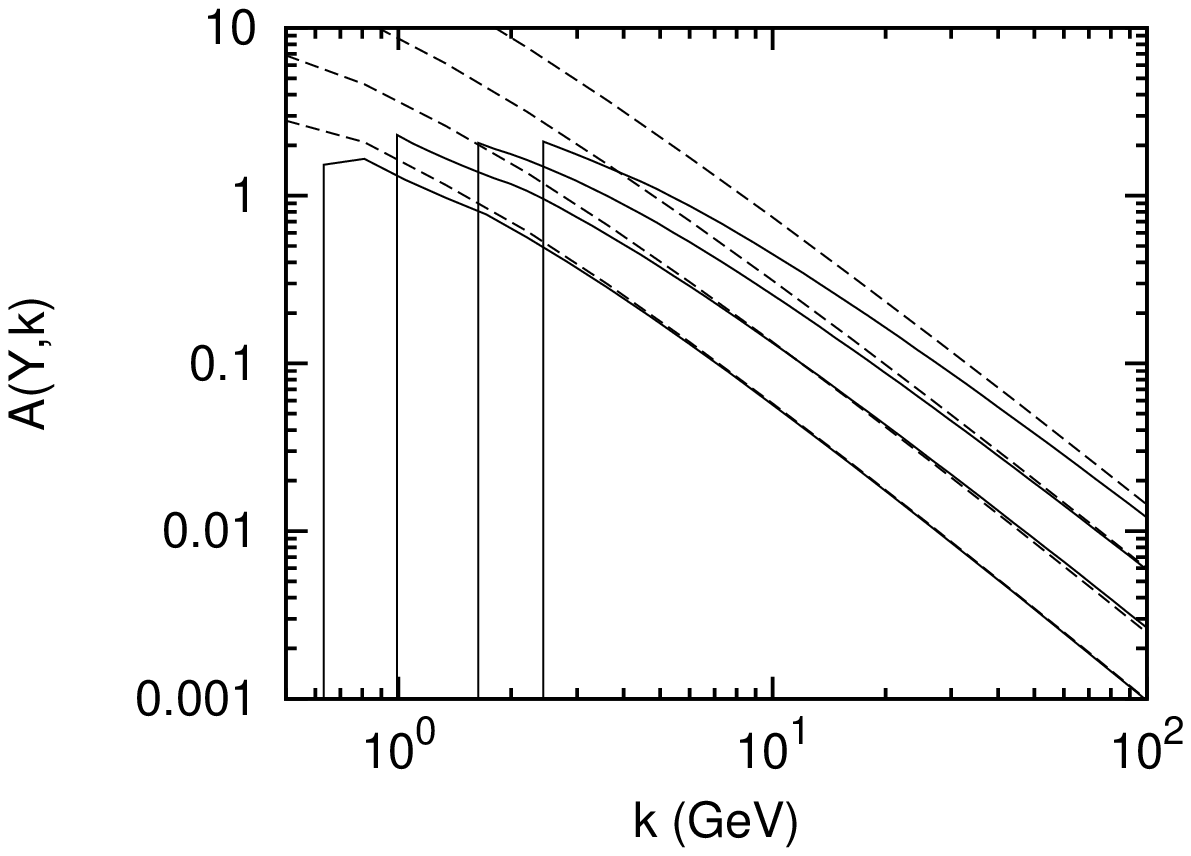}
  \includegraphics[angle=270, scale=0.6]{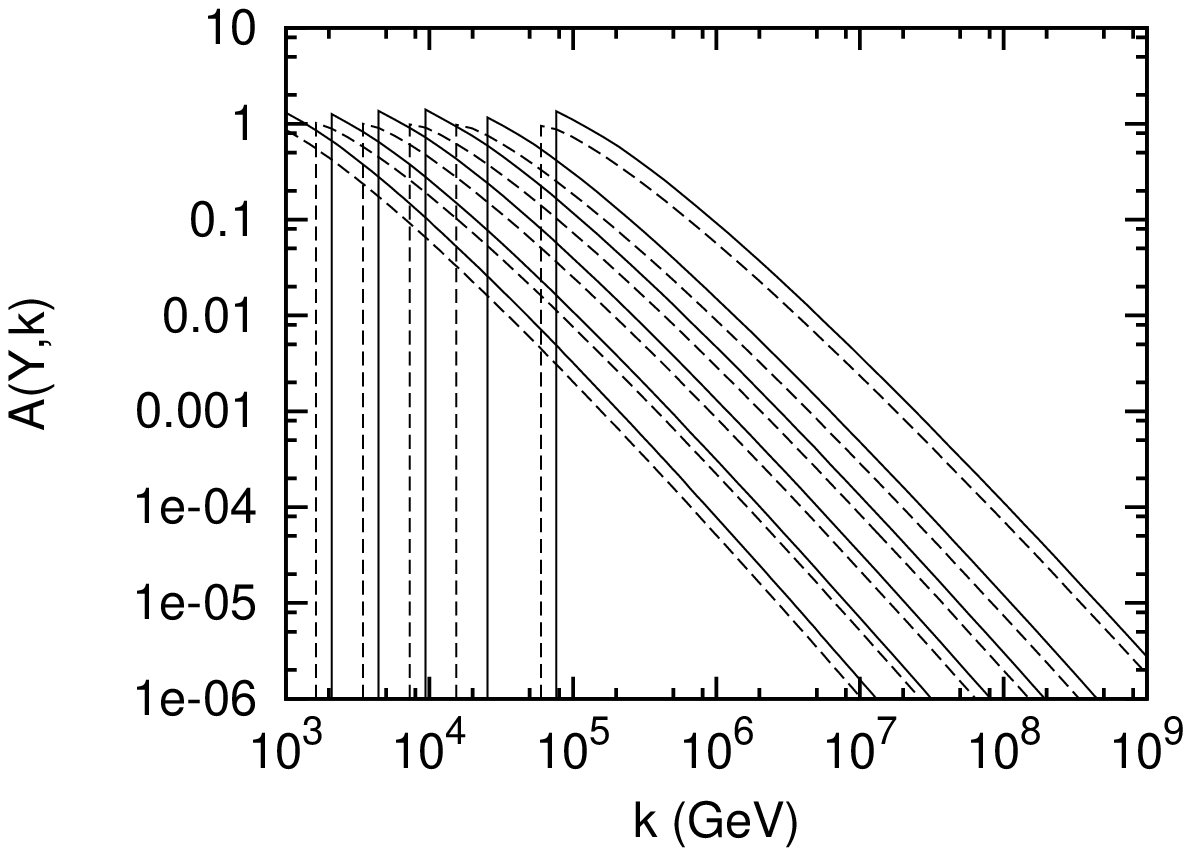}}
\caption {\sl Left: Solutions to the running--coupling CCFM equation
\eqref{eq:ccfmdiffeq3} with the saturation boundary  (solid lines)
  and without it (dashed lines) for
  $Y=8, \,10,\, 12,$ and $14$. Right: CCFM  (solid lines) vs.
   BFKL (dashed lines) solutions with running coupling
   and saturation boundary for very high rapidities:
   $Y= 60, \,70,\,80,\,90,\,100,$ and $120$.
  \label{fig:bfklres6} }}
  \end{figure}

It is first interesting to compare the predictions of
\eqnum{eq:ccfmdiffeq3} to those of the BFKL equation for the strictly
linear evolution. This is shown in Fig.~\ref{fig:bfklres4} for both fixed
and running coupling, and one clearly sees that the BFKL evolution is
considerably faster. This difference is to be attributed to the
non--local term in the r.h.s. of \eqnum{eq:ccfmdiffeq3}: the `retarded'
distribution $\mathcal{A}(Y - \log(k'^2/k^2), k')$ is generally smaller
than the `instantaneous' one $\mathcal{A}(Y, k')$.

But even though the CCFM evolution is somewhat slower,
\eqnum{eq:ccfmdiffeq3} still shows a pronounced growth with $Y$, which in
the absence of any non--linearity would rapidly lead to unitarity
violation. To cure for that, we now enforce the absorptive boundary
condition \eqref{BC} on \eqnum{eq:ccfmdiffeq3}. The corresponding results
are compared to those of the purely (CCFM) linear evolution in
Fig.~\ref{fig:bfklres5}, for both fixed and running coupling. As in the
BFKL case (compare to Fig.~\ref{fig:bfklrunres1}), the difference is more
pronounced for a running coupling, since then the linear evolution is
again infrared--unstable, and this instability is removed by the
inclusion of saturation.

For the more realistic, running--coupling, case it is furthermore
interesting to show the results at lower rapidities, as relevant for LHC.
This is exhibited in the leftmost figure in Fig.~\ref{fig:bfklres6},
together with the corresponding results of the strictly linear evolution.
As one can see there, for $Y=14$ and $k$ as high as 10 GeV (which is well
above the respective saturation momentum $Q_s\simeq 2.5$~GeV),
saturation reduces the predicted gluon distribution by about a factor of
2.

Finally, in the rightmost figure in Fig.~\ref{fig:bfklres6}, we compare
the saturation fronts provided by the BFKL and CCFM evolutions with
running coupling and for relatively high rapidities, up to $Y=120$. What
is remarkable about this last figure is that the BFKL and CCFM evolutions
with saturation and running coupling appear to very close to each other
and for all rapidities --- meaning that the corresponding fronts
propagate at roughly the same speed. This is at variance with the
corresponding situation at {\em fixed} coupling, where the BFKL evolution
is still faster. This is confirmed by the estimates of the saturation
momentum for the CCFM evolution, as extracted from fits to our numerical
results: It is numerically clear that $\rho_s$ rises linearly with $Y$
for a fixed coupling, and it exhibits a $\sqrt{Y}$ behaviour for a
running coupling, just like for BFKL. Trying simple fits of the form
$\rho_s = \lambda_f \abar Y$ and, respectively, $\rho_s = \lambda_r
\sqrt{Y}$, we find the values\footnote{Note that, at fixed coupling, the
CCFM estimate for the parameter $\lambda_f$ defined as above is still a
function of $\abar$ (unlike the respective BFKL estimate); hence this
value $\lambda_f \approx 3.5$ must be seen as the value corresponding to
$\abar=0.2$. We shall further discuss the $\abar$--dependence of the CCFM
parameter $\lambda_f$ in Ref.~\citep{Avsar}.} $\lambda_f \approx 3.5$
and, respectively, $\lambda_r \approx 2.8$. For fixed coupling, this
value $\lambda_f$ for the saturation exponent is indeed smaller (although
not {\em much} smaller) than the corresponding BFKL estimate
$\lambda_f\simeq 4.9$. But for running coupling, the CCFM and BFKL
estimates for $\lambda_r$ are essentially the same within our numerical
accuracy. Moreover, the manifest similarity between the shapes of the
BFKL and CCFM fronts in Fig.~\ref{fig:bfklres6} (right) suggests that the
CCFM fronts exhibit the same properties of geometric scaling as the BFKL
ones (again, for a running coupling).

It would be of course interesting to understand this similarity between
the BFKL and CCFM evolutions towards saturation in more depth, and also
to perform more detailed studies of the CCFM evolution with saturation
boundary, in order e.g. to explicitly test geometric scaling. We postpone
such studies to a further work \citep{Avsar}.

\subsection*{Acknowledgments}
We would like to thank Al Mueller and Dionysis Triantafyllopoulos for
valuable comments on the manuscript. This work is supported in part by
Agence Nationale de la Recherche via the programme ANR-06-BLAN-0285-01.


\providecommand{\href}[2]{#2}\begingroup\raggedright\endgroup

\end{document}